\documentclass[final,3p,times,onecolumn]{elsarticle}
\usepackage{amsmath,amssymb,amsfonts,scalerel}
\usepackage{blindtext}
\usepackage{graphicx}
\usepackage{adjustbox}
\usepackage{epsfig}
\usepackage{float}
\usepackage{subfig}
\usepackage{xcolor}
\usepackage[colorlinks=true,citecolor=blue,linkcolor=blue,urlcolor=blue]{hyperref}
\usepackage{setspace}
\usepackage{multirow}
\usepackage{multicol}
\usepackage{url}
\usepackage{siunitx}
\usepackage{hyperref}
\usepackage{rotating}
\usepackage{geometry}
\biboptions{sort,compress}
\journal{Chemical Physics Letters}

\begin{document}

\begin{frontmatter}

\title{Molecular dynamics simulation of ion dynamics within 
       SPEEK polymer electrolyte of PEM Fuel Cells}

\author{S.S.~Awulachew$^{1}$}
\ead{sshinie21@gmail.com}
\author[]{K.N.~Nigussa$^{2}$\corref{cor1}}
\cortext[cor1]{Corresponding author:\ kenate.nemera@aau.edu.et\ 
               (K.N.~Nigussa)}

\address{$^1$Department of Physics,\ Haramaya University,\ P.O. Box 138,\ 
             Dire Dawa,\ Ethiopia\\
         $^2$Department of Physics,\ Addis Ababa University,\ P.O. Box 1176,\ 
             Addis Ababa,\ Ethiopia}
         
\begin{abstract}
Proton transport\ property is studied\ 
by modelling\ the intermolecular pair\ 
correlation\ functions\ of\ the proton\ 
ion\ with the electrode and\ the electrolyte\ 
of a polymer electrolyte fuel cell~(PEMFC)\ 
by using Materials-Studio\ and\ then applying\ 
molecular dynamics\ simulation.\   
A\ stable\ structure of\ the novel\ 
electrode\ design\ is\ obtained using\ 
density functional\ theory.\ When\ 
the polymer\ electrolyte\ is assumed\ 
as\ anhydrous,\ the efficiency\ of the\ 
proton\ transport\ increases.\  
Analysis of the\ proton\ coordination\ 
numbers shows that\ more protons are\ 
found\ in\ the region of oxygen\ than\ 
sulfur atoms\ of the Sulfonic acid\ Ether\ 
Ester\ Ketone~(SEEK)\ electrolyte.\ The\ 
proton\ conductivity values\ are increased\ 
with including\ interaction\ effects\ 
from\ electrode compared to without.\ 
At a temperature\ of 333 K,\ these\ 
values of ion\ conductivity are\ 
$7.69 \times 10^{5} \hspace{1mm}\mathrm{S \hspace{1mm}cm^{-1}}$\ 
and 
$4.28 \times 10^{5} \hspace{1mm}\mathrm{S \hspace{1mm}cm^{-1}}$,\
respectively,\ with and without.\ The\ 
network of\ the hydrogen\ bond\ is the\ 
path\ to transport\ of protons\ via\ the\ 
processes\ of hydrogen bond\ creating\ 
and\ breaking.\ The\ values\ of\ highest\ 
peaks\ in the radial\ distribution\ function\ 
($g(r)$)\ calculations\ appear\ to\ fall\ 
in the hydrogen\ bond\ formation\ region.\ 
Thus,\ it looks\ that a combination of\ 
Pd$\rm_{3}$Ag\ as an electrode and unhydrated\ 
SEEK as an electrolyte\ could make up for\ 
a cost effective\ components\ of new\ design\ 
PEMFC.\   
\end{abstract}


\end{frontmatter}

\section{Introduction\label{sec:intro}}
Low temperature\ polymer electrolyte membrane\ 
fuel cells\ (PEMFC),\ also known as proton exchange\ 
membrane fuel cells,\ have stimulated interest in\ 
recent years as\ a feasible power source,\ 
particularly in\ automotive applications.\
PEM fuel cells\ generate electricity by combining\ 
hydrogen\ (or hydrogen-rich fuel) and oxygen\ 
(from the air),\ resulting in low pollution\ 
and high fuel economy~\cite{vielstich2009handbook}.\
The development of\ highly active and low-cost catalysts\ 
is a major challenge\ before proton exchange membrane\
fuel cells (PEMFCs)\ can find large-scale practical\ 
applications~\cite{wieckowski2003catalysis,weber2018carbon,
antolini2003formation, shao2007understanding,yu2007recent, 
chen2009shape, peng2009designer, liu2011tuning}.

Platinum has\ been widely chosen\ for a catalytic\ 
role in\ PEMFC since a\ decades ago~\cite{HL69}.\ 
However,\ there is\ a need to find\ alternative\ 
material\ to replace Pt,\ since it has\ been\ 
used as a\ catalyst for fuel\ cells for\ a long\ 
time\ and\ might be\ in short\ supply in the\ 
future.\ One\ of the\ major barriers\ to\ 
commercialization\ of\ fuel cell engines\ is\ 
the electrodes'\ high cost,\ which is\ mostly\ 
due to\ the\ use of expensive\ platinum~(Pt)-containing\ 
electro-catalysts.\
In addition,\ platinum-based\ catalysts can\ be\ 
poisoned by\ carbon monoxide~(CO)\ and other\ 
pollutants~\cite{shao2006palladium}.\ 
Because of these\ characteristics,\ pursuing for\ 
alternative\ catalyst materials\ have\ 
gotten a\ lot of research\ attention.\ 
The oxygen reduction\ reaction~(ORR)\ 
at the cathode in\ the PEM fuel cell\
is a\ rate-limiting phase,\ in which molecular\ 
oxygen\ is\ dissociated and\ mixed with protons\ 
and\ electrons\ supplied by\ the anode through\ 
the membrane and\ external circuit\ to form water\ 
as\ by-product.\ 

\begin{figure}[htbp!]
\centering
\begin{adjustbox}{max size ={\textwidth}{\textheight}}
\includegraphics[scale=0.2]{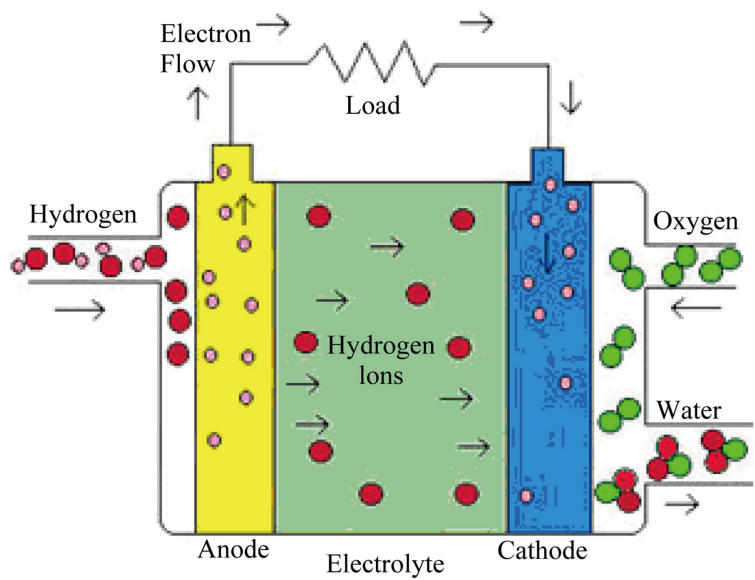}
\end{adjustbox}
\caption{Simplified diagram of the PEM fuel cell\ 
         operating process,\ taken from~\cite{shao2006palladium}.
         \label{fig1}}
\end{figure}

A\ recent\ study\ by\ K.N.~Nigussa~\cite{Nigussa_2019}\
has\ investigated\ palladium\ as\ a possible\ 
candidate\ for\ replacement.\ This\ corresponds\ to\ 
$\textit{\rm{x}}=0$,\ in\ the\ alloy\ composition.\ 
However,\ still\ because\ of\ the\ rare\ availability\ 
of\ Pd,\ consideration\ of\ further\ other\ 
alloys\ as\ a\ replacement\ candidate\ is\ of\ 
interest\ in\ this\ work.\ 
As a result,\ one of our primary\ motives is to\ 
investigate the properties\ of the Pd$\rm_{1-x}$Ag$\rm_{{x}}$\ 
($\textit{\rm{x}}=0.25-0.75$)\ alloy as\ components\ 
of fuel\ cell as electrode.\ This\ could\ also\ 
represent\ a\ study\ on\ a\ part\ of\ a\ grand\ alloy\ 
matrix\ Pt$\rm_{1-x}$Ag$\rm_{x}$Pd$\rm_{1-x}$Ag$\rm_{x}$,\
making\ the\ fuel\ cell\ electrode,\ where\ investigating\ 
how\ certain\ region\ of\ the\ alloy\ dominated\ by\ 
Pd$\rm_{1-x}$Ag$\rm_{x}$\ behaves\ is\ the\ central\ 
interest\ point\ of\ this\ work.\ Studying\ efficiency\ 
of\ candidate\ replacement\ catalysts\ goes\ through\ 
investigating\ interactions\ of\ molecules\ 
of\ fuel\ with\ components\ of\ the\ fuel cell\  
as\ well\ as\ also\ transport\ of\ the\ molecules\ 
of\ fuel\ within\ the\ fuel\ cell.\   
The ion dynamics in\ the electrolyte\ of\ 
the\ fuel cell\ is studied by\ Molecular Dynamics~(MD).\ 
Thus,\ this\ activity\ is\ a\ conjugate\ study\ 
to\ our\ preceding\ study~\cite{awulachew2022principles}\ 
on\ analysis\ of\ the\ property\ of\ Pd$\rm_{3}$Ag~(111)\ 
as\ a\ potential\ electrode\ component\ of\ a\ 
fuel\ cell.\ A simplified diagram\ of the PEM fuel 
cell\ operating principles\ is shown in Fig.~\ref{fig1}.\
%.........Figure 1.............................

\section{Computational Methods\label{sec:comp}}
\subsection{Force Field}
\begin{figure}[ht!]
\centering
\begin{adjustbox}{max size ={\textwidth}{\textheight}}
\includegraphics[scale=0.7]{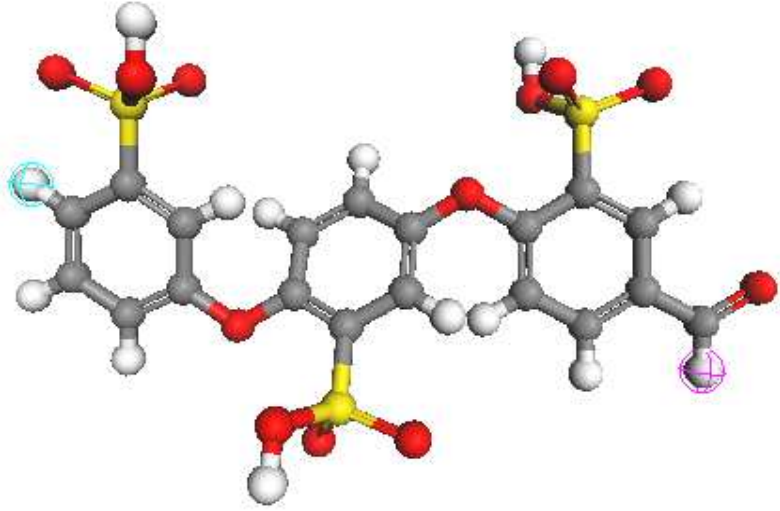}
\end{adjustbox}
\caption{Sulfonic acid Ether Ester Ketone~(SEEK).
Color online.~Colors:~O-red,\ S-yellow,\ 
C-dark~grey,~and~H-white.\label{fig2}}    
\end{figure}

\begin{figure}[ht!]
\centering
\begin{adjustbox}{max size ={\textwidth}{\textheight}}
\includegraphics[scale = 0.4]{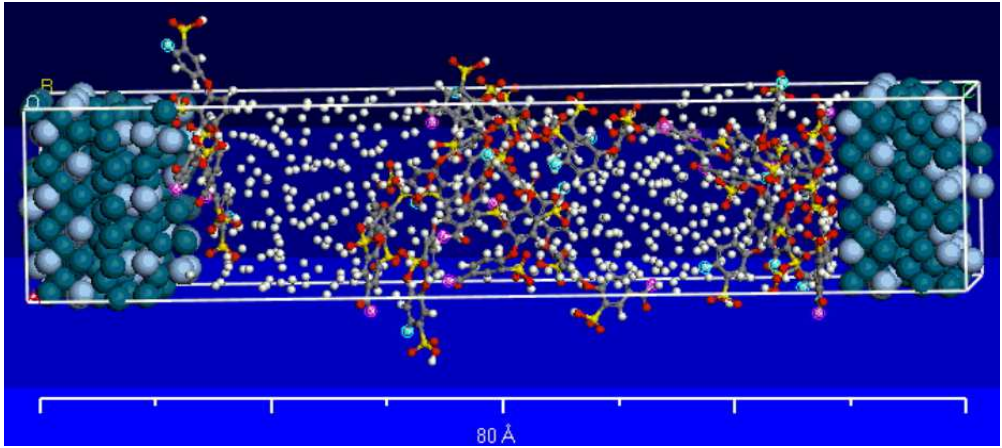}
\end{adjustbox}
\caption{Fuel Cell\ Model for\ Hydrogen ion\ mobility\
with Pd$\rm_{3}$Ag electrode and SEEK electrolyte.\
Color online.~Colors:~O-red,\ S-yellow, C-dark~grey,\ 
H-white,\ Ag-cyan,~and~Pd-light~grey.\label{fig3}}   
\end{figure}

\begin{figure}[ht!]
\centering
\begin{adjustbox}{max size ={\textwidth}{\textheight}}
\includegraphics[scale = 0.7]{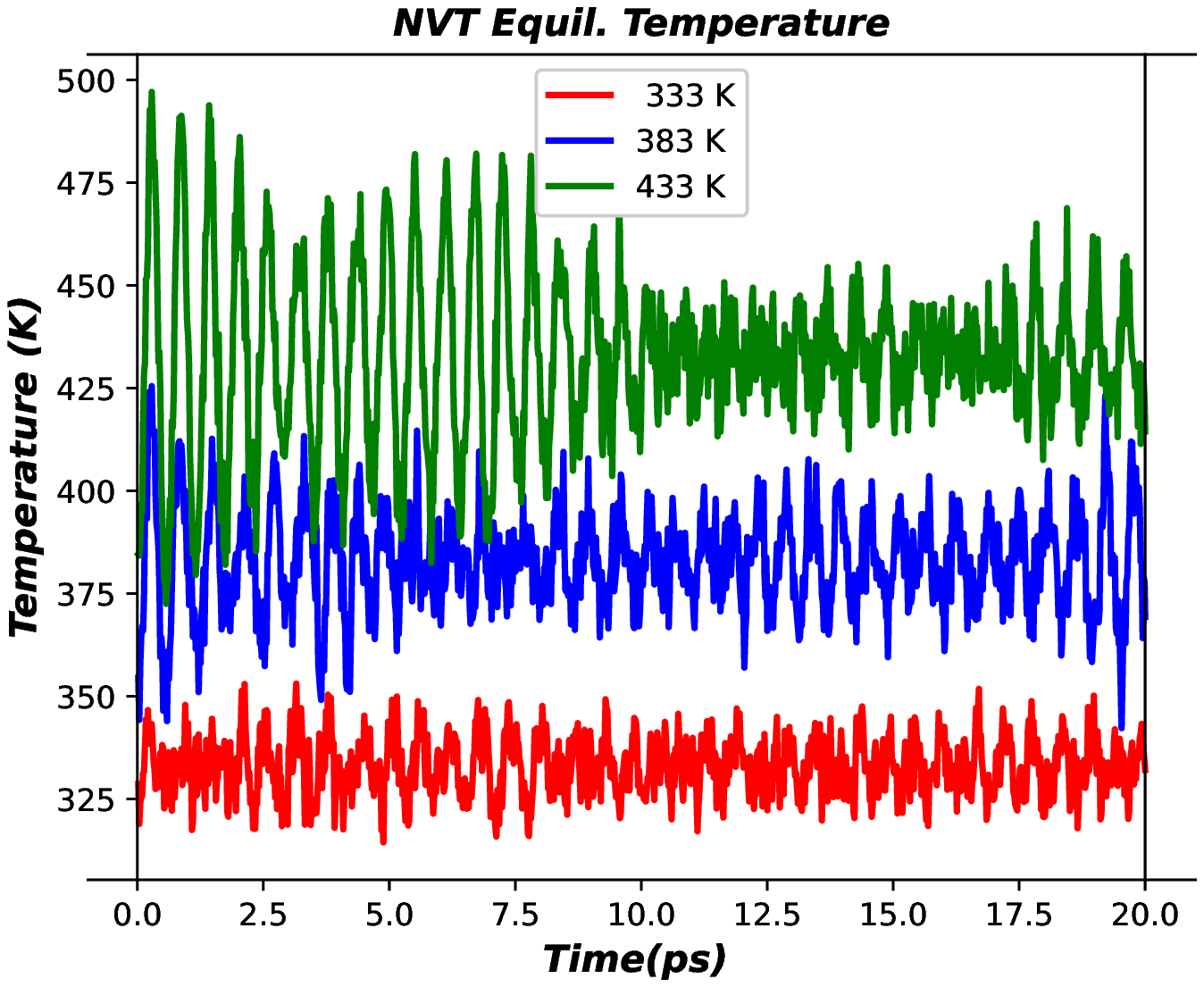}
\end{adjustbox}
\caption{Temperature variation\ of the simulation\ 
         system\ during NVT equilibration.\label{fig4}}
\end{figure}

\begin{figure}[ht!]
\centering
\begin{adjustbox}{max size ={\textwidth}{\textheight}}
\includegraphics[scale =0.7]{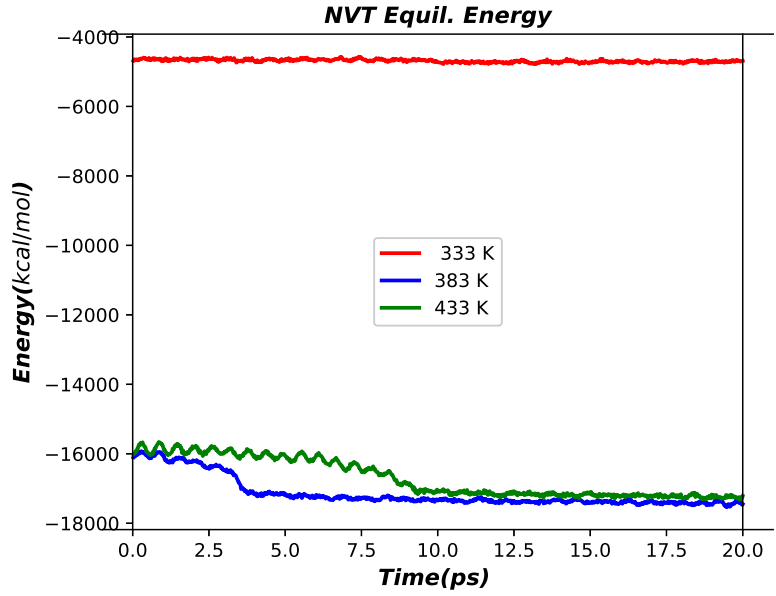}
\end{adjustbox}
\caption{Total potential\ energy variation\ of\ 
         the simulation\ system during NVT\ 
         equilibration.\label{fig5}}
\end{figure}

\begin{figure}[ht!]
\centering
\begin{adjustbox}{max size ={\textwidth}{\textheight}}
\includegraphics[scale = 0.7]{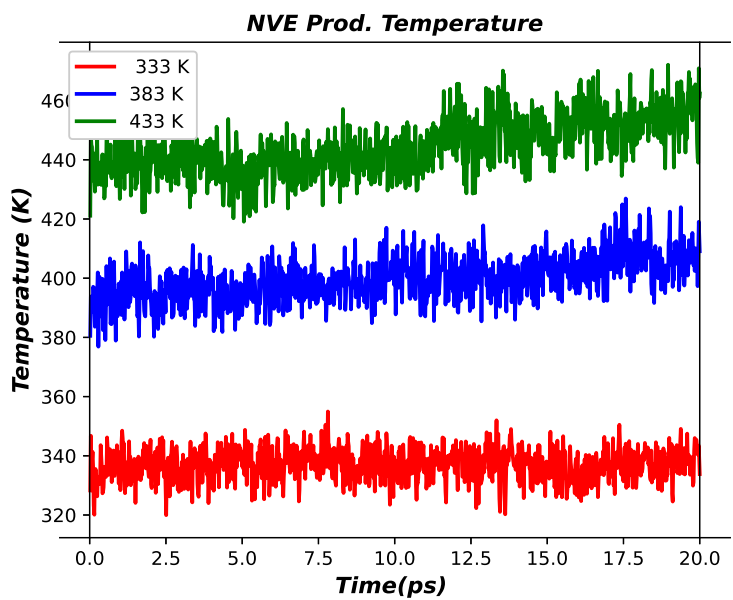}
\end{adjustbox}
\caption{Temperature variation\ of the simulation\ 
         system\ during NVE~production run.
         \label{fig6}}    
\end{figure}

\begin{figure}[ht!]
\centering
\begin{adjustbox}{max size ={\textwidth}{\textheight}}
\includegraphics[scale = 0.7]{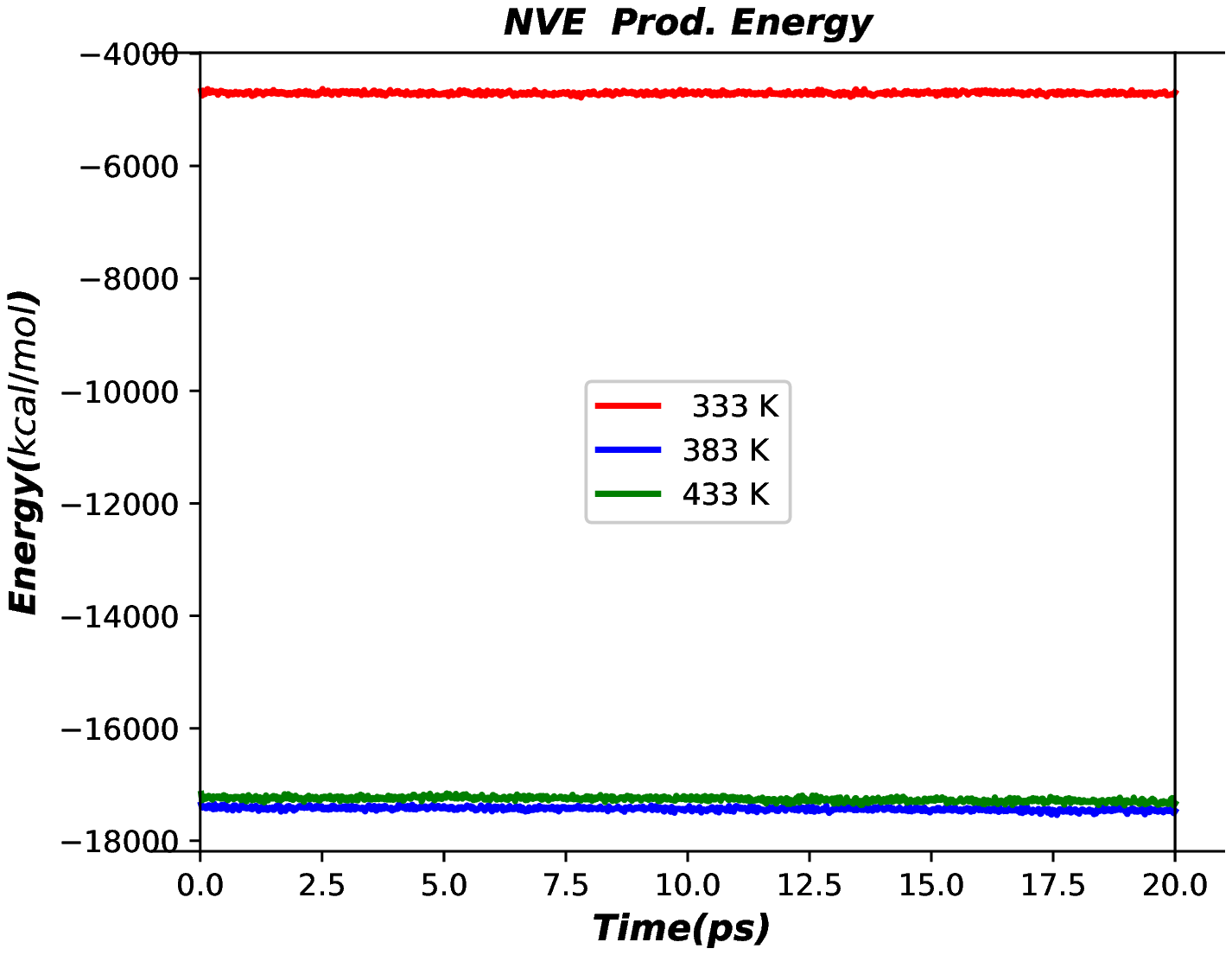}
\end{adjustbox}
\caption{Total potential\ energy variation\ of the\ 
         simulation\ system during\ NVE~production\ 
         run.\label{fig7}}
\end{figure}

We\ assume\ a\ simplified\ model\ of\ PEMFCs.\ 
For this work,\ Sulfonic acid\ 
Ether Ester\ Ketone~(SEEK) Polymer\ with\ chemical\
formula\ unit\ of~(C$\rm_{19}$H$\rm_{14}$O$\rm_{12}$S$\rm_3$)\ 
was\ used\ as electrolyte to\ investigate\ the proton conductivity,\ 
diffusion\ constant,\ radial distribution function,\
and\ coordination number,\ by developing\ a model\ 
as shown\ in Figs.~\ref{fig2}~$\&$~\ref{fig3}.\ 
%%%%%...............Figure 2................
%%%%...................Figure 3.................
The Amorphous Cell~(AC)\ for\ simulation was\ 
constructed\ with\ 400 protons\ (H$^+$),\ and 15\ 
Sulfonated\ Ether\ Ester\ Ketone~(SEEK)\ polymer\ 
chain.\ The AC\ was\ placed\ between\ layers\ of\ 
anode~and~cathode\ which\ are\ made\ of\ 
Pd$\rm_{3}$Ag~(111)\ alloy~(see Fig.~\ref{fig3}).\  
At\ initial\ stages,\ the simulation cell\ was\ 
constructed\ by inserting\ the molecules\ 
randomly.~The simulation\ system consists\ 
of H, C, S, O, Pd, and Ag atoms.\ The COMPASS\ 
force field is\ generated\ according\ to\ a\ 
literature~\cite{sun1998compass},\ with\ 
a three-dimensional\ periodic boundary\ 
conditions\ being\ applied.\ A "smart" method\ 
was\ used to\ optimize the geometry\ of the\ 
cell.\ The final\ structures\ obtained from\ 
the\ geometry\ optimization\ process\ were heated\ 
to\ temperatures\ ranging from 200 K to 500 K.\
The annealing process\ was carried out five\ 
times.\ The dynamic simulation is\ done\ using\ 
the final structures\ created from\ the annealing\ 
procedure.\ The dynamic properties\ of these\ 
cells were\ studied using a 20 ps\ NVE production\ 
run following a 20 ps NVT\ equilibration.\
The Nose-Hoover\ approach was\ used to\ manage\ 
temperature during\ the simulation process.\ 
The time step\ was 1 fs\ and the setting of the\ 
temperature\ was~(333 K,~383 K,~433 K).\ The\ 
cut-off radius\ of the non-bonded\ interactions\ 
was set\ to 12.5~{\AA}.\ Ewald summation was\
employed for\ both van der\ Waals and Coulombic\ 
interactions~\cite{nymand2000ewald, ibrahim2020go}.\ 
The total potential\ energy and\ temperature\ 
variations of\ the simulation systems\ during NVT\ 
equilibration\ and NVE~production run\ are shown\ 
in Figs.~\ref{fig4}-\ref{fig7}. 
%%%%%.............Figure 4....................

%%%%...............Figure 5...................
%%%%%%%............Figure 6...................

%%...............Figure 7.......................

As shown\ in the\ figures,\ the fluctuations\ 
were\ small\ and are\ in the normal\ range\ 
which\ can be seen to\ confirm\ occurrence\ 
of\ NVT\ equilibration.\ The temperature\ 
fluctuations\ and almost the\ unchanged\ 
energy may\ be caused\ by the continuous\ 
transformation between\ kinetic\ energy\ 
and potential energy.\ For all cases,\ 
the volume of\ the simulation\ box is\ 
23206.05~$\rm{\AA}^3$.\ In\ the\ case\ 
of\ Pt\ and Pd electrodes,\ the\ Pd$\rm_{3}$Ag~(111)\
in Fig.~\ref{fig3}\ is\ replaced with\ 
Pt~(111)\ and\ Pd~(111),\ respectively.\\ 
\subsection{Molecular Dynamics~(MD)~simulation}
In the MD simulation,\ the mobile trajectory\ 
of proton\ can generate\ mean square\ 
displacement~(MSD),\ and\ the average\ 
square of displacement,\ where a\ 
particle moves\ from time 0\ to time $t$.\
The self-diffusion\ coefficient is directly\ 
related\ to the time\ of\ correlation of\ 
coordinates.\ The\ self-diffusion\ 
coefficients~$D$\ is obtained\ from\ 
the slope of mean\ square displacement~(MSD)\ 
as a\ function of\ time $t$ according\ 
to the Einstein\ relation~(see~Eq.~\eqref{eq1}\ 
and\ a\ literature~\cite{cummings1991nonequilibrium}).\  
\begin{equation}{\label{eq1}}
D =\frac{1}{6}\lim_{t\rightarrow \infty}\frac{dMSD(t)}{dt} 
\end{equation}
where MSD is defined as 
\begin{equation}{\label{eq2}}
MSD(t) = \langle \frac{1}{N}\sum_{i=1}^{N}|\mathbf{r}_i(t)
-\mathbf{r}_0(0)|^2\rangle
\end{equation}
where $\mathbf{r}_i(t)$ is the position of\ 
the atom\ $i$\ at\ the time $t$, and the\ 
$\langle{\cdots}\rangle$\ denotes\ an\ 
average\ over all the\ time steps.\ 
N represents\ the\ number of diffusible particles.\
The proton conductivity~($\sigma_{cond}$,\ in units\ 
of\ Siemens per\ centimeter~([\textit{S/cm}]))\ is\ 
calculated\ using\ the diffusion\ coefficient of\ 
proton\ which\ is\ provided by\ Eq.~\eqref{eq3},\  
\begin{equation}{\label{eq3}}
{\sigma}_{cond} = \frac{N{z^2}{e^2}}{V{k\rm_{B}}T}D\rm_{proton}
\end{equation}
in which\ $N$ represents\ the number of protons,\ 
$z$ is\ the charge on the proton~(+1 in this study),\ 
$e$\ the elementary charge~($1.6{\times}10^{-19}$ C),\ 
$D\rm_{proton}$\ is\ the diffusion\ coefficient of\ 
proton,\ $V$ the\ volume of the\ simulation cell,\ 
$k\rm_{B}$\ is\ the Boltzmann's\ 
constant~($1.38 {\times} 10^{-23}$ J/K),\ and $T$\ 
the absolute temperature.\\
Radial distribution\ function~($g(r)$)\ characterizes\ 
the\ local structure\ at short-range.\ 
It reflects the local\ structural\ arrangements\ 
around\ any given atom\ in a short distance\ and\ 
describes\ how distribution\ density\ varies as a\ 
function\ of distance.\
%%%....................Figure 8............
\begin{figure}[htbp!]
\centering
\begin{adjustbox}{max size ={\textwidth}{\textheight}}
\includegraphics[scale = 0.7]{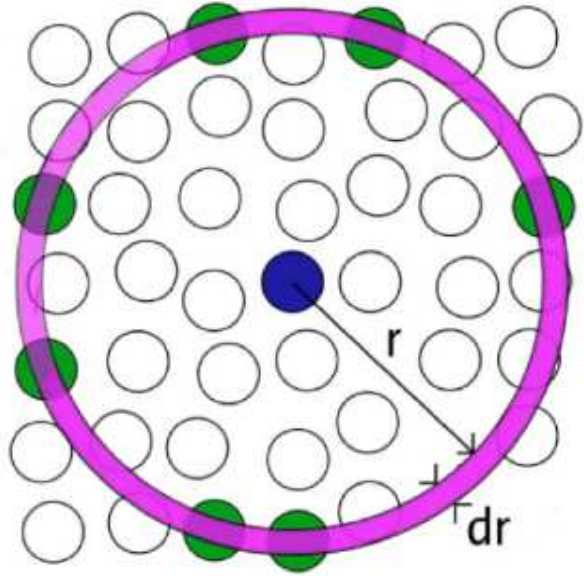}
\end{adjustbox}
\caption{Space discretization\ for the evaluation\ 
        of\ the radial\ distribution function,\ taken\ 
        from~\cite{allen1987analyse}.\label{fig8}}

\end{figure}

For a homogeneous\ atomic system,\ the radial\
distribution\ function is\ shown in\ 
Fig.~\ref{fig8}~(taken\ from~\cite{allen1987analyse}),\ 
where the blue\ particle is a\ reference particle\ 
at the\ origin O\ and green particles\ are those\ 
which\ are within\ the pink\ circular shell.\ 
If $\rho = \frac{N}{V}$\ is\ the average\ number\ 
density\ of particles,\ then the\ local\ 
time-averaged\ density at\ a distance\ $r$\ 
from O\ is ${\rho} g(r)$.\ In other words,\ 
$g(r)$\ involves\ determining how\ many particles\ 
are\ within a distance\ of $r$ and $r + dr$\ away\ 
from\ the blue\ particle at\ the origin.\ $g(r)$\ 
is usually\ normalized as\ the ratio\ between the\ 
number\ of local particles\ and the average\ 
number of particles in\ the system.\
The normalized\ $g(r)$ also represents\ the\ 
probability\ of finding\ an atom in\ a shell\ 
$dr$ at a distance\ $r$ from a\ reference point\ 
atom.\ Thus,\ $g(r)$ can\ be used to\ calculate\ 
the number\ of atoms in the\ shell $dr$\
\begin{equation}
\text{d} N(r) = \frac{N}{V}g(r)4{\pi}{r^2}dr 
\label{coordinationeq}
\end{equation}
where $N(r)$ denotes\ the number of atoms\ 
as a\ function of location,\ $N$ the total\ 
number of atoms,\ and $V$\ is\ the system's\ 
volume.\ When there are\ multiple chemical\ 
species,\ the\ partial\ radial distribution\ 
function\ $g\rm_{{\alpha}{\beta}}(r)$\ can be\ 
calculated\ by\ 
\begin{equation}
g_{\alpha\beta}(r) = \frac{dN_{\alpha\beta}(r)}
{4\pi {r^2}dr \rho_{\alpha}} 
\label{rdfeq}
\end{equation}
were $\rho_{\alpha}=\frac{N_{\alpha}}{V}$\ denotes\ 
the\ density of\ $\alpha$ species,\ and $N_{\alpha}$\ 
denotes\ the total number of $\alpha$ species in\ 
the system.\
%%%%%%%%%%%%%%%%%%%%%%%%%%%%%%%%%%%%%%%%%%%%%%
\section{Results and Discussion \label{sec:res}}
%%%%%Table 1...........................
\begin{table*}[htbp!]
\addtolength{\tabcolsep}{1.0mm}
\renewcommand{\arraystretch}{1.3}
\centering
\small
\caption{The maxima of the\ radial distribution functions\ 
        in~[\AA], conductivity in~[S/cm]\ and coordination\ 
        numbers of the\ O-H and S-H\ pair at different\ 
        temperatures in~[K].~The\ first two\ 
        rows at 333~K~(No eld.)~for O-H\ and S-H is\ 
        without\ electrode\ effect included\ 
        in\ the\ interaction\ pair functions.\label{tab1}}
\begin{tabular}{lccccc}
\hline
{Electrode} & {Temperature} & {Correlation Pair} & {Highest Peak} & {Conductivity} & Coordination Number\\ 
\hline
\multirow{12}*{Pt} & {333} & O-H & 2.35 & $1.44\times 10^{6}$ & 38.24\\
           {}&{}&S-H& 4.33& $1.44\times 10^{6}$ & 34.24 \\
            \cline{3-6}
           {} & {383} & O-H & 2.35 & $1.41\times 10^{6}$ & 38.12\\
           {} & {}    & S-H & 4.59 & $1.41\times 10^{6}$ & 34.12\\
           \cline{3-6}
           {} & {433} & O-H & 2.41 & $1.24\times 10^{6}$ & 36.47\\ 
           {} & {} & S-H & 4.59 & $1.24\times 10^{6}$ & 32.49\\ 
           \cline{2-6}
           {} & {200} & O-H & 2.35 & $1.85\times 10^{6}$ & 40.60\\
           {} & {} & S-H & 4.61 & $1.85\times 10^{6}$ & 37.28\\
            \cline{3-6}
           {} & {250} & O-H & 2.35 & $1.73\times 10^{6}$ & 39.14\\
           {} & {} & S-H & 4.61 & $1.73\times 10^{6}$ & 35.83\\
            \cline{3-6}
           {} & {300} & O-H & 2.37 & $1.72\times 10^{6}$ & 38.54\\
           {} & {} & S-H & 4.23 & $1.72\times 10^{6}$ & 35.00\\
           \cline{2-6}

\multirow{12}*{Pd} & {333} & O-H & 2.37 & $1.41\times 10^{6}$ & 38.62\\
           {}&{}&S-H&4.57 & $1.41\times 10^{6}$& 34.73\\
            \cline{3-6}
           {} & {383} & O-H & 2.35 & $1.24\times 10^{6}$ & 38.11\\
           {} & {}    & S-H & 4.37 & $1.24\times 10^{6}$ & 34.00\\
           \cline{3-6}
           {} & {433} & O-H & 2.37 & $1.31\times 10^{6}$ & 38.32\\ 
           {} & {} & S-H & 4.57 & $1.31\times 10^{6}$ & 33.80\\ 
           \cline{2-6}
           {} & {200} & O-H & 2.35 & $1.59\times 10^{6}$ & 38.48\\
           {} & {} & S-H & 4.59 & $1.59\times 10^{6}$ & 34.44\\
            \cline{3-6}
           {} & {250} & O-H & 2.33 & $2.01\times 10^{6}$ & 41.28\\
           {} & {} & S-H & 4.57 & $2.01\times 10^{6}$ & 36.99\\
            \cline{3-6}
           {} & {300} & O-H & 2.37 & $1.81\times 10^{6}$ & 40.64\\
           {} & {} & S-H & 4.57 & $1.81\times 10^{6}$ & 36.00\\
           \cline{2-6}

\multirow{12}*{Pd$\rm_{3}$Ag} & {333~(No electrode)} & O-H & 2.37 & $4.28\times 10^{5}$ & 40.72\\
           {}&{}&S-H&4.42 &$4.28\times 10^{5}$ & 37.43 \\
            \cline{3-6}
           {} & {383} & O-H & 2.35 & $7.69\times 10^{5}$ & 34.48\\
           {} & {}    & S-H & 4.37 & $7.69\times 10^{5}$ & 30.53\\
           \cline{3-6} 
           {} & {383} & O-H & 2.37 & $7.56\times 10^{5}$ & 37.17\\
           {} & {}    & S-H & 4.77 & $7.56\times 10^{5}$ & 33.82\\
           \cline{3-6}
           {} & {433} & O-H & 2.37 & $7.46\times 10^{5}$ & 35.27\\ 
           {} & {} & S-H & 4.23 & $7.46\times 10^{5}$ & 31.01\\ 
           \cline{2-6}
           {} & {200} & O-H &2.33  &$9.37\times 10^{5}$ & 35.10\\
           {} & {} & S-H & 4.45 & $9.37\times 10^{5}$ & 31.10\\
            \cline{3-6}
           {} & {250} & O-H & 2.33 & $9.09\times 10^{5}$ & 35.98\\
           {} & {} & S-H & 4.41 & $9.09\times 10^{5}$ & 31.93\\
            \cline{3-6}
           {} & {300} & O-H & 2.35 & $7.33\times 10^{5}$ & 33.91\\
           {} & {} & S-H & 4.63 & $7.33\times 10^{5}$ & 29.94\\
           \cline{2-6}                    
\hline
\end{tabular}
\end{table*}
\subsection{Dynamics properties of protons}
%%%%%%%%%%%%%%%%%%%%%
\subsubsection{Pt as an Electrode}
The dynamic\ properties\ of protons can be\ 
determined\ using the mean\ square displacement~(MSD)\ 
of the\ protons during\ the simulation.\
%%%%Figure 9....................................
\begin{figure}[htbp!]
\centering
\begin{adjustbox}{max size ={\textwidth}{\textheight}}
\includegraphics[scale=0.7]{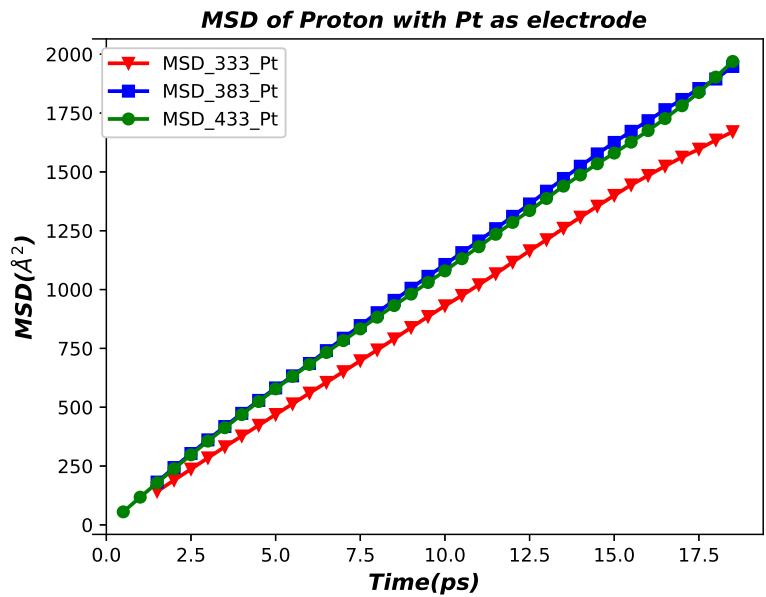}
\end{adjustbox}
\caption{The mean square\ displacement~(MSD) of\ 
         protons.~The simulations\ were\ carried\ 
         out\ at 333 K,\ 383 K,\ and 433 K\ with\ 
         Pt as electrode,\ over a period of\ 
         20 ps.\label{fig9}}
\end{figure}
As\ shown in Fig.~\ref{fig9},~MSD of protons at\ 
333 K\ is\ slightly increased\ in\ the\ presence\ 
of\ electrode\ interaction.\ At 333 K,\ the\ 
self-diffusion\ coefficient\ of proton\ is\ 
$1.52 \times 10^{-3}\hspace{1mm}\mathrm{cm^2 s^{-1}}$,\ 
at 383 K,\ it is\ $1.73 \times 10^{-3} 
\hspace{1mm}\mathrm{cm^2 s^{-1}}$,\ 
and~at 433 K,\ it is $1.72 \times 10^{-3} 
\hspace{1mm}\mathrm{cm^2 s^{-1}}$ as shown in\ 
Table~\ref{tab1}.\

\begin{figure}[htbp!]
\centering
\begin{adjustbox}{max size ={\textwidth}{\textheight}}
\includegraphics[scale=0.7]{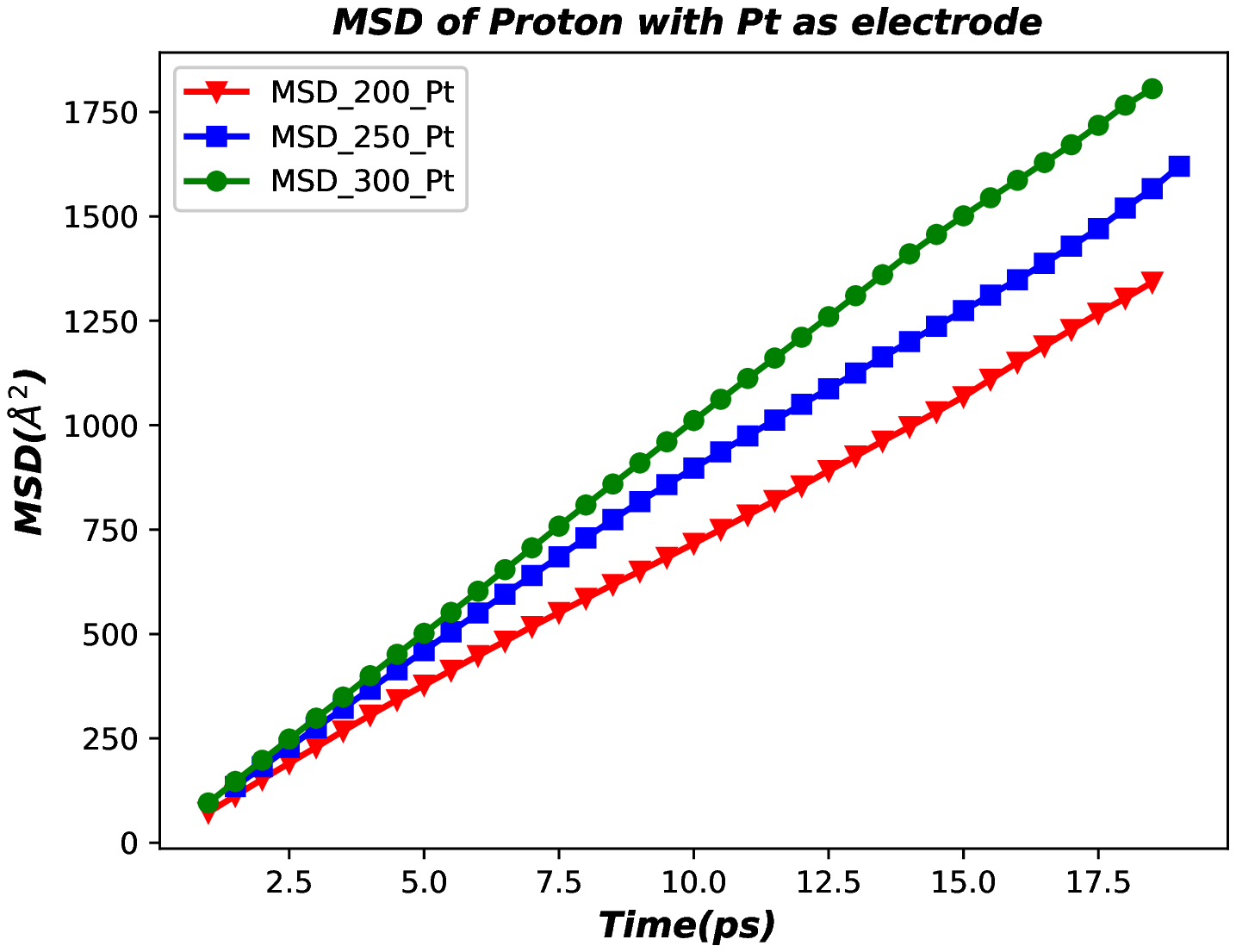}
\end{adjustbox}
\caption{The mean square displacement~(MSD) of\ 
         protons.~The simulations\ were\ carried out\ 
         at 200 K,~250 K,~and 300 K, with Pt electrode\
         over a period\ of 20 ps.~The\ electrode\ 
         interaction\ effect\ is\ included.\label{fig10}}
\end{figure}

\begin{figure}[htbp!]
\centering
\begin{adjustbox}{max size ={\textwidth}{\textheight}}
\includegraphics[scale = 0.7]{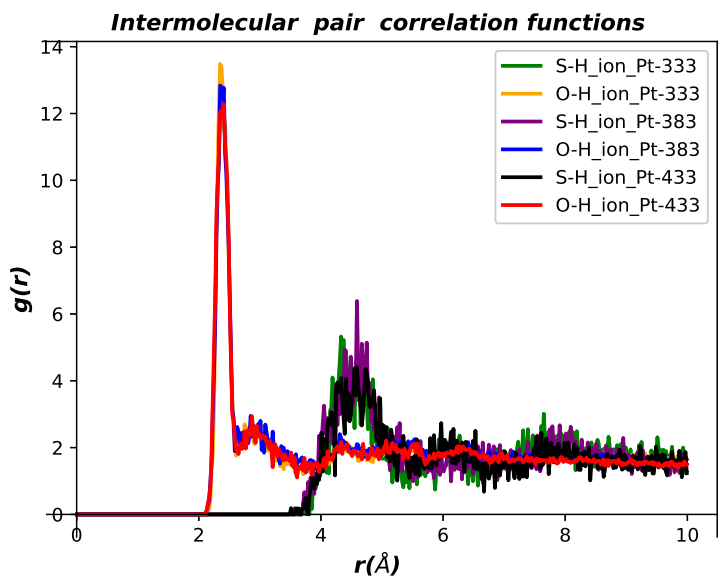}
\end{adjustbox}
\caption{Intermolecular\ pair correlation\ functions of\ 
         O-H and S-H\ at 333 K,\ 383 K,\ and 433 K~with\ 
         Pt electrode.\label{fig11}}
\end{figure}

It\ looks\ that\ the\ MSD\ value\ increases\ as\ 
temperature\ increases\ within\ the\ temperature\ 
range\ considered.~The results\ were obtained\ 
using\ Eq.~\eqref{eq2}.~The\ MSD\ plots\ at\ 
lower\ temperatures\ are\ shown\ in\ Fig.~\ref{fig10}.\ 
The self-diffusion\ coefficient at temperatures\ 
of\ 200 K,\ 250 K,\ and 250 K,~where\ 
interaction\ effects\ from electrode\ 
included\ are~$1.18 \times 10^{-3} \hspace{1mm}\mathrm{cm^2 s^{-1}}$,\ 
$1.38 \times 10^{-3} \hspace{1mm}\mathrm{cm^2 s^{-1}}$ and\ 
$1.65 \times 10^{-3} \hspace{1mm}\mathrm{cm^2 s^{-1}}$,\ 
respectively,\ as shown in Table~\ref{tab1}.\
%............Figure 10..............
The proton conductivity is\ calculated by using\ 
Eq.~\eqref{eq3}.~The\ outcomes\ are\
$1.44 \times 10^{6} \hspace{1mm}\mathrm{S \hspace{1mm}cm^{-1}}$,\
at 333 K,\ $1.41 \times 10^{6} \hspace{1mm}\mathrm{S \hspace{1mm}cm^{-1}}$\ 
at 383 K\ and $1.24 \times 10^{6}\hspace{1mm}\mathrm{S \hspace{1mm}cm^{-1}}$\ 
at 433 K as shown in Table~\ref{tab1}.\ 
Furthermore,\ the computational results\ 
suggests that the\ conductivity\ decreases with\ 
temperature\ increases.\ The proton conductivity\ 
corresponding\ to\ temperatures\ of\ 200 K,~250 K,\
and 300 K,~are\ 
$1.85 \times 10^{6} \hspace{1mm}\mathrm{S \hspace{1mm}cm^{-1}}$,\
$1.73 \times 10^{6} \hspace{1mm}\mathrm{S \hspace{1mm}cm^{-1}}$,~and\ 
$1.72 \times 10^{6}\hspace{1mm}\mathrm{S \hspace{1mm}cm^{-1}}$,\ 
respectively as shown in Table~\ref{tab1}.
Intermolecular pair\ correlation functions of O-H pair,\ 
S-H pair,\ at 333 K,\ 383 K and 433 K \ with\ Pt\ 
as an\ electrode are presented\ in Fig.~\ref{fig11}.\ 
%%%Figure 11...............................

\begin{figure}[htbp!]
\centering
\begin{adjustbox}{max size ={\textwidth}{\textheight}}
\includegraphics[scale=0.7]{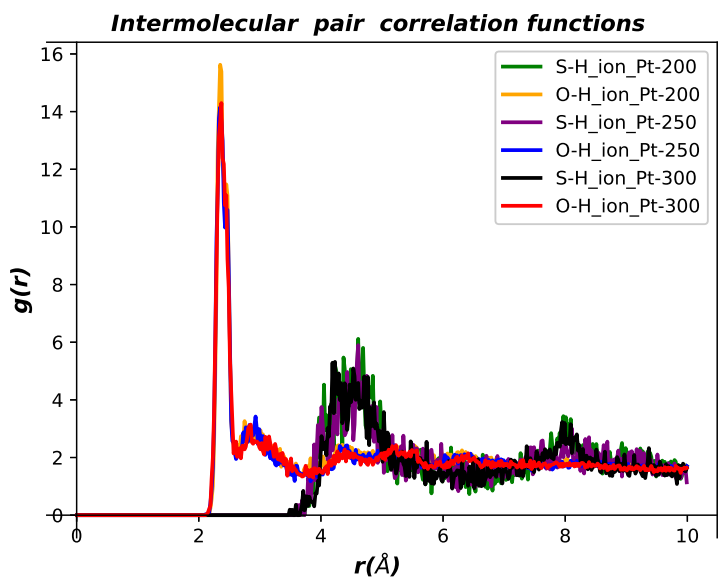}
\end{adjustbox}
\caption{Intermolecular pair correlation functions of O-H 
         and S-H at 200 K, 250 K and 300 K~Pt electrode.\
         Here\ the\ number\ of\ protons is\ 
         400.\label{fig12}}
\end{figure}

The highest peaks\ and coordination\ number\ of O-H\ 
and\ S-H pair\ are \ presented\ in\ Table~\ref{tab1}.\
All O-H pair\ highest\ peak distances\ falls well\ 
with\ in the range of\ a hydrogen bond.\ 
The\ intermolecular\ pair correlation functions\ 
of O-H pair,~S-H pair,~at 200 K,~250 K,~and 300 K 
are shown in Fig.~\ref{fig12}.\ The highest peaks\ 
of O-H pair intermolecular pair correlation functions\ 
occur\ at a positions of~2.35~{\AA},~2.35~{\AA},\ 
and 2.37~{\AA},~for 200 K,~250 K,~and~300 K,~respectively.\ 
The coordination number of O-H pair are 41,~39,~and 38,\ 
respectively,\ for 200 K,~250 K,~and 300 K.~The highest\ 
peaks of S-H pair\ intermolecular pair correlation\ 
functions are\ at a positions of~4.61~{\AA},\ 4.61~{\AA},\ 
and 4.23~{\AA},~respectively,~for 200 K,~250 K,~and\ 
300 K.\ The coordination number\ of S-H pair are\ 
37,\ 36,\ and 35,\ respectively,\ for 200 K,\ 250 K,\ 
and,\ 300 K~(as shown in Table~\ref{tab1}).\ 
The\ fractional coordination\ numbers are results\ 
of a selected\ numerical integrator\ of the Eq.~\eqref{coordinationeq},\
and\ a corresponding\ nearest\ value integers\ 
can be used\ upon\ discussions.\ 

All O-H pair\ highest peak\ distances fall well\ 
within the\ range of a hydrogen\ bond formation\ 
region. 
 %%%%%%%Figure 12.........................
According to the values of highest peaks falls\ 
in hydrogen bond formation region,\ oxygen\ 
interacts with\ protons by creating hydrogen\ 
bonds,\ but sulfur\ interacts with protons\ 
at a greater\ distance than oxygen.\ The\ 
hydrogen bonding\ network can be\ used\ 
to transfer protons\ from anode to cathode.\
\subsubsection{Pd as an Electrode}
%%%%Figure 13...............................
\begin{figure}[htbp!]
\centering
\begin{adjustbox}{max size ={\textwidth}{\textheight}}
\includegraphics[scale=0.7]{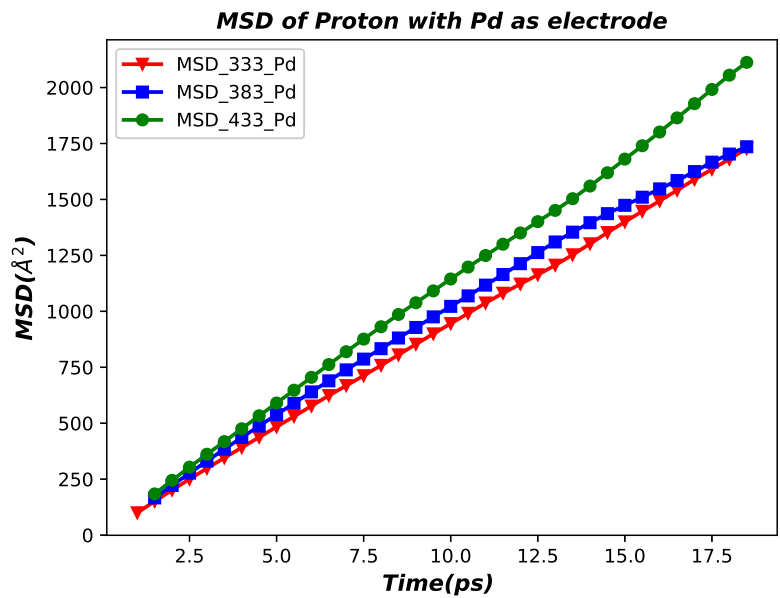}
\end{adjustbox}
\caption{The mean square displacement~(MSD) of\ 
         protons.~The simulations\ were\ carried out\ 
         at 333 K,\ 383 K and 433 K\ with Pd as\  
         electrode,\ over a period of 20 ps.\
         \label{fig13}}
\end{figure}
\begin{figure}[htbp!]
\centering
\begin{adjustbox}{max size ={\textwidth}{\textheight}}
\includegraphics[scale=0.7]{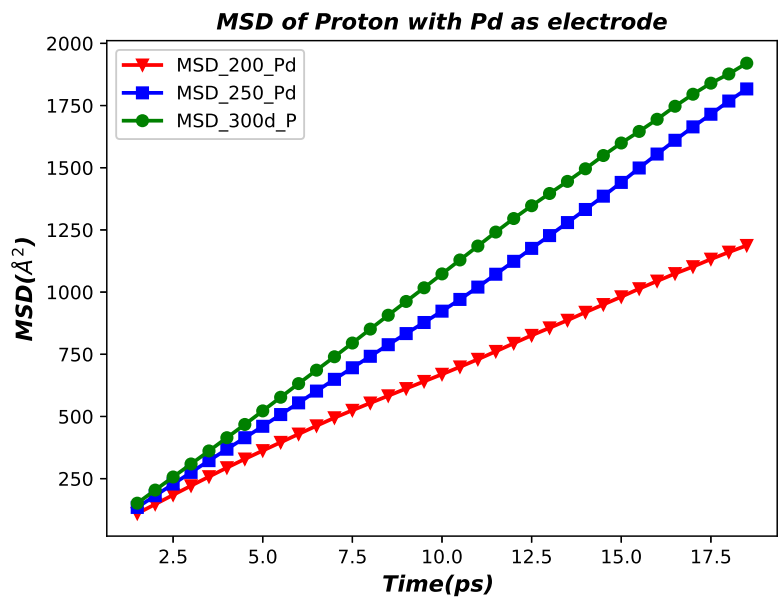}
\end{adjustbox}
\caption{The mean square displacement~(MSD) of\ 
         protons.~The simulations\ were\ carried out\ 
         at 200 K,~250 K,~and 300 K, with Pd electrode\ 
         over a period\ of 20 ps.~The\ electrode\ 
         interaction\ effect\ is\ included.\label{fig14}}
\end{figure}

As\ shown in Fig.~\ref{fig13},~MSD of protons\ 
at 333 K\ is\ slightly increased\ in\ the\ 
presence\ of\ electrode\ interaction.\ At 333 K,\ 
the self-diffusion coefficient\ of proton\ is\ 
$1.53 \times 10^{-3} \hspace{1mm}\mathrm{cm^2 s^{-1}}$,\ 
at 383 K\ it is\ $1.55 \times 10^{-3} \hspace{1mm}\mathrm{cm^2 s^{-1}}$,\ 
and~at 433 K\ it is $1.85 \times 10^{-3} 
\hspace{1mm}\mathrm{cm^2 s^{-1}}$ as shown\ 
in Table~\ref{tab1}.~It\ looks\ that\ 
the\ MSD\ value\ increases\ as\ temperature\ 
increases\ within\ the\ temperature\ range\ 
considered.~The results\ were obtained\ using\ 
Eq.~\eqref{eq2}.~The\ MSD\ plots\ at\ lower\ 
temperatures\ are\ shown\ in\ Fig.~\ref{fig14}.\ 
The self-diffusion coefficient at temperatures\ 
of\ 200 K,\ 250 K,\ and 250 K,~where\ 
interaction\ effects\ from electrode\ 
are\ included\ are~$1.04 \times 10^{-3} \hspace{1mm}\mathrm{cm^2 s^{-1}}$,\ 
$1.64 \times 10^{-3} \hspace{1mm}\mathrm{cm^2 s^{-1}}$,\ 
and\ 
$1.77 \times 10^{-3} \hspace{1mm}\mathrm{cm^2 s^{-1}}$,\ 
respectively as shown in Table~\ref{tab1}.\
%%%%%Figure 14...............................

\begin{figure}[htbp!]
\centering
\begin{adjustbox}{max size ={\textwidth}{\textheight}}
\includegraphics[scale = 0.7]{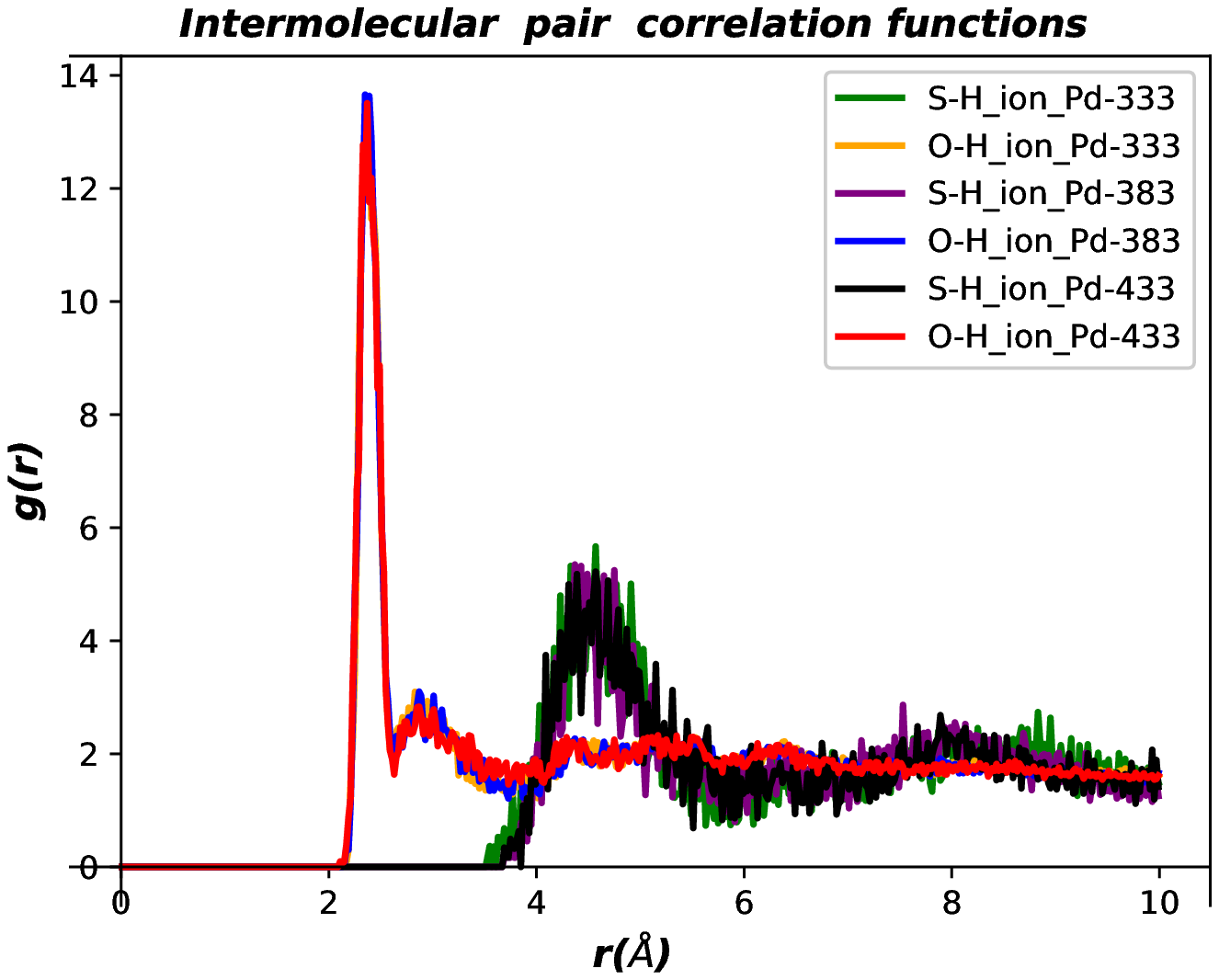}
\end{adjustbox}
\caption{Intermolecular\ pair correlation\ functions of\ 
         O-H and S-H\ at 333 K,\ 383 K,\ and  433 K~with\ 
         Pd electrode.\label{fig15}}
\end{figure}

The proton conductivity is\ calculated by using\ 
Eq.~\eqref{eq3}.~The\ outcomes\ are\
$1.41 \times 10^{6} \hspace{1mm}\mathrm{S \hspace{1mm}cm^{-1}}$\
at 333 K, $1.24 \times 10^{6} \hspace{1mm}\mathrm{S \hspace{1mm}cm^{-1}}$\ 
at 383 K\ and $1.31 \times 10^{6}\hspace{1mm}\mathrm{S \hspace{1mm}cm^{-1}}$\ 
at 433~K as shown in Table~\ref{tab1}. 
Furthermore,\ the computational results\ suggest\ 
that the\ conductivity\ decreases with temperature\ 
increases.\ The proton conductivity corresponding\ to\
temperatures\ of\ 200 K,~250 K,~and 300 K,~are\ 
$1.59 \times 10^{6} \hspace{1mm}\mathrm{S \hspace{1mm}cm^{-1}}$,\
$2.01 \times 10^{6} \hspace{1mm}\mathrm{S \hspace{1mm}cm^{-1}}$,~and\ 
$1.81 \times 10^{6}\hspace{1mm}\mathrm{S \hspace{1mm}cm^{-1}}$,\ 
respectively as shown in Table~\ref{tab1}.
Intermolecular pair\ correlation functions of O-H pair,\ 
S-H pair,\ at 333 K, 383 K and 433 K \ with\ Pd as an 
electrode are presented\ in Fig.~\ref{fig15}.\ 
%%%Figure 15...................................

The highest peaks\ and coordination\ number\ of O-H and\ 
S-H pair\ are \ presented\ in\ Table~\ref{tab1}.~All\ 
O-H pair highest\ peak distances\ falls well within\ 
the range of a hydrogen bond.\ 

\begin{figure}[htbp!]
\centering

\begin{adjustbox}{max size ={\textwidth}{\textheight}}
\includegraphics[scale=0.7]{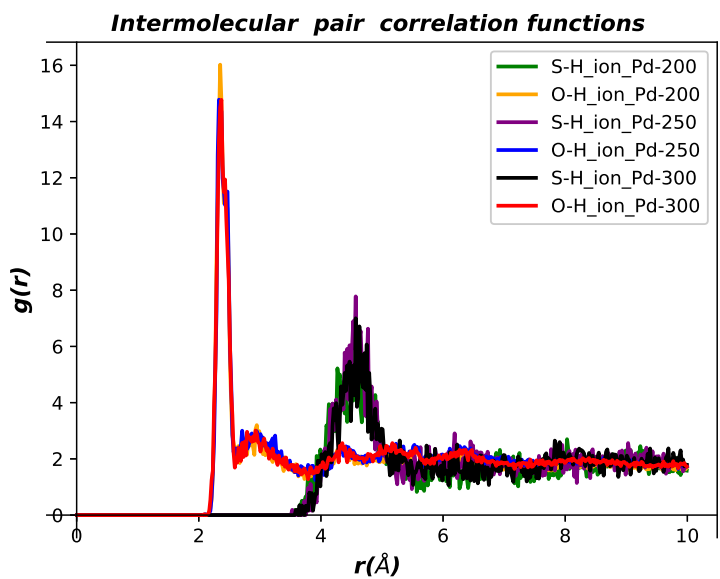}
\end{adjustbox}
\caption{Intermolecular pair correlation functions of O-H and S-H  
         at 200 K, 250 K and 300 K with Pd electrode.~Here\ the\ 
         number\ of\ protons is 400.\label{fig16}}
\end{figure}

The\ intermolecular pair correlation functions\ 
of O-H pair,~S-H pair,~at 200 K,~250 K,~and 300 K 
are shown in Fig.~\ref{fig16}.\ The highest peaks\ 
of O-H pair intermolecular pair correlation functions\ 
occur\ at a positions of~2.35~{\AA},~2.35~{\AA},\ 
and 2.37~{\AA},~for 200 K,~250 K,~and~300 K,~respectively.\ 
The coordination number of O-H pair are 41,~39,~and 38,\ 
respectively,\ for 200 K,~250 K,~and 300 K.~The highest\ 
peaks of S-H pair\ intermolecular pair correlation\ 
functions are\ at a positions of~4.61~{\AA},\ 4.61~{\AA},\ 
and 4.23~{\AA},~respectively,~for 200 K,~250 K,~and\ 
300 K.\ The coordination number\ of S-H pair are\ 
37,\ 36,\ and 35,\ respectively,\ for 200 K,\ 250 K,\ 
and,\ 300 K~(as shown in Table~\ref{tab1}).\ 
The\ fractional coordination\ numbers are results\ 
of a selected\ numerical integrator\ of the Eq.~\eqref{coordinationeq},\
and\ a corresponding\ nearest\ value integers\ 
can be used\ upon\ discussions.\ 

All O-H pair\ highest peak\ distances fall well\ 
within the\ range of a hydrogen\ bond formation\ 
region. 
 %%%%%%%Figure 16............................
According to the values of highest peaks falls\ 
in hydrogen bond formation region,\ oxygen\ 
interacts with\ protons by creating hydrogen\ 
bonds,\ but sulfur\ interacts with protons\ 
at a greater\ distance than oxygen.
\subsubsection{Pd$\rm_{3}$Ag as an Electrode}
%%%%Figure 17...........................
\begin{figure}[htbp!]
\centering
\begin{adjustbox}{max size ={\textwidth}{\textheight}}
\includegraphics[scale=0.7]{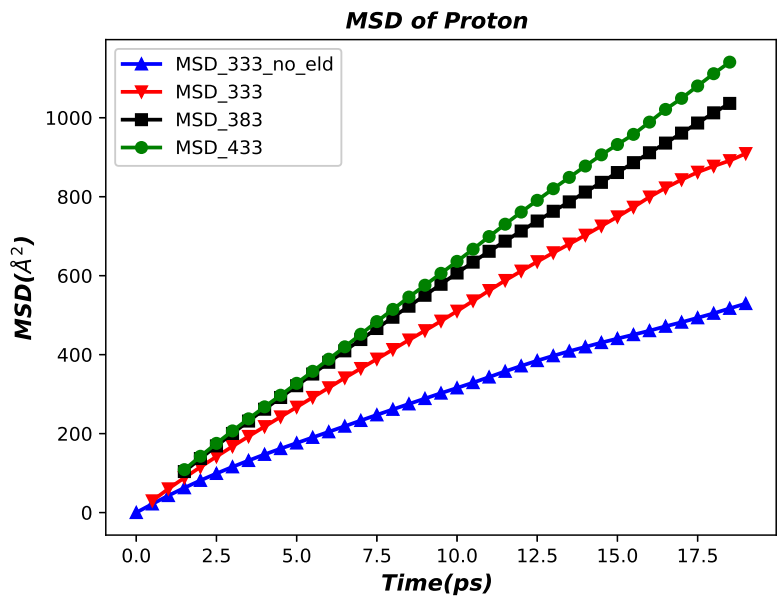}
\end{adjustbox}
\caption{The mean square displacement~(MSD) of\ 
         protons.~The simulations\ were\ carried out\ 
         at 333 K with and without Pd$\rm_{3}$Ag\ 
         electrode, at\ 383 K and 433 K with\ Pd$\rm_{3}$Ag\ 
         electrode,\ over a period\ of 20 ps.\label{fig17}}
\end{figure}
\begin{figure}[htbp!]
\centering
\begin{adjustbox}{max size ={\textwidth}{\textheight}}
\includegraphics[scale=0.7]{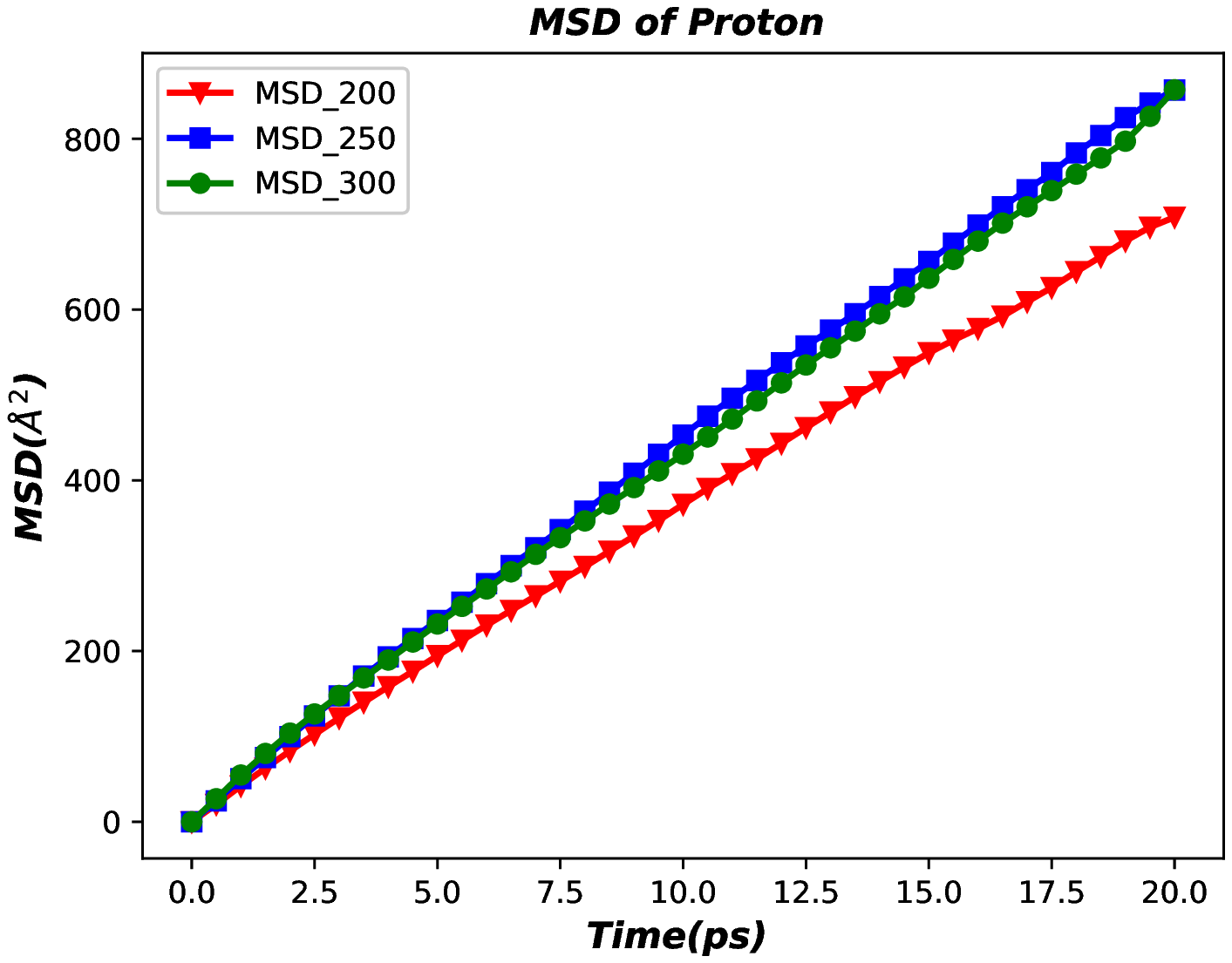}
\end{adjustbox}
\caption{The mean square displacement~(MSD) of\ 
         protons.~The simulations\ were\ carried out\ 
         at 200 K,~250 K,~and 300 K,~with\ Pd$\rm_{3}$Ag\ 
         electrode,\ over a period\ of 20 ps.~The\ 
         electrode\ interaction\ effect\ is\ 
         included.\label{fig18}}
\end{figure}

As\ shown in Fig.~\ref{fig17},~MSD of protons at\ 
333 K\ is\ increased\ in\ the\ presence\ of\ 
electrode\ interaction\ than without electrode.\ 
So,\ thus,\ at 383 K,\ and 433 K, the simulations\ 
are done by including\ interaction\ effects from\ 
electrode.\ At 333 K with electrode,\ the self-diffusion\ 
coefficient\ of proton\ is\ 
$8.01 \times 10^{-4} \hspace{1mm}\mathrm{cm^2 s^{-1}}$,\ 
while without\ electrode\ it is $4.46 \times 10^{-4} \hspace{1mm} 
\hspace{1mm}\mathrm{cm^2 s^{-1}}$,\ at 383 K\ with electrode\ 
it is\ $9.06 \times 10^{-4} \hspace{1mm}\mathrm{cm^2 s^{-1}}$,\ 
and~at 433 K with electrode it is $1.01 \times 10^{-3} 
\hspace{1mm}\mathrm{cm^2 s^{-1}}$ as shown in Table~\ref{tab1}.\
It\ looks\ that\ the\ MSD\ value\ increases\ as\ 
temperature\ increases\ within\ the\ temperature\ 
range\ considered.~The results\ were obtained\ using\ 
Eq.~\eqref{eq2}.~The\ MSD\ plots\ at\ lower\ 
temperatures\ as\ shown\ in\ Fig.~\ref{fig18}.\ 
The self-diffusion coefficient at temperatures\ 
of\ 200 K,\ 250 K,\ and 250 K,~where\ interaction\ 
effects\ from electrode\ are\ included\ 
are~$5.86 \times 10^{-4} \hspace{1mm}\mathrm{cm^2 s^{-1}}$,\ 
$7.11 \times 10^{-4} \hspace{1mm}\mathrm{cm^2 s^{-1}}$ and\ 
$6.88 \times 10^{-4} \hspace{1mm}\mathrm{cm^2 s^{-1}}$,\ 
respectively as\ shown in Table~\ref{tab1}.\
%%%%%Figure 18.................................

\begin{figure}[htbp!]
\centering
\begin{adjustbox}{max size ={\textwidth}{\textheight}}
\includegraphics[scale = 0.7]{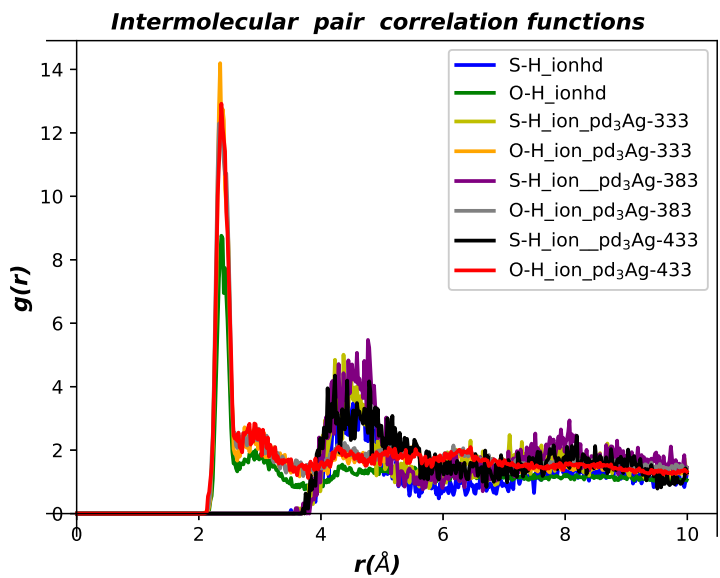}
\end{adjustbox}
\caption{Intermolecular\ pair correlation\ functions of\ 
         O-H and S-H\ at 333 K with and without Pd$\rm_{3}$Ag\ 
         electrode,\ at 383 K and 433 K~with\ Pd$\rm_{3}$Ag\
         electrode.\label{fig19}}

\end{figure}

The proton conductivity is\ calculated by using\ 
Eq.~\eqref{eq3}.~The\ outcomes\ are\
$4.28 \times 10^{5} \hspace{1mm}\mathrm{S \hspace{1mm}cm^{-1}}$,\
and~$7.69 \times 10^{5} \hspace{1mm}\mathrm{S \hspace{1mm}cm^{-1}}$,\
respectively,\ without\ and with electrode,~at 333 K.\
At\ temperatures\ of 383 K~and~433 K and in\ the\ 
presence\ of\ interaction\ effect\ from\ electrode,\ 
the values of\ conductivity\ are\ 
$7.56 \times 10^{5}\hspace{1mm}\mathrm{S \hspace{1mm}cm^{-1}}$,\ 
and~$7.46 \times 10^{5}\hspace{1mm} \mathrm{S \hspace{1mm}cm^{-1}}$,\ 
respectively as shown in Table~\ref{tab1}.~Furthermore,\ 
the computational results\ suggests that the\ conductivity\ 
decreases as temperature\ increases.\ The proton\ 
conductivity corresponding\ to\ temperatures\ of\ 
200 K,~250 K,~and 300 K,~are\ 
$9.37 \times 10^{5} \hspace{1mm}\mathrm{S \hspace{1mm}cm^{-1}}$,\
$9.09 \times 10^{5} \hspace{1mm}\mathrm{S \hspace{1mm}cm^{-1}}$,~and\ 
$7.33 \times 10^{5}\hspace{1mm}\mathrm{S \hspace{1mm}cm^{-1}}$,\ 
respectively as shown in Table~\ref{tab1}.
Intermolecular pair\ correlation functions of O-H pair,\ 
S-H pair,\ at different temperature\ without and with\ 
electrode are illustrated\ in Fig.~\ref{fig19}.\ 
%%%Figure 19..............................

The highest peaks\ and coordination\ number\ of O-H and\ 
S-H pair\ by\ adding water\ to the system\ without\ 
electrode\ are calculated\ with values 2.37~{\AA}\ 
and~36.49 for O-H pair,\ and 4.41~{\AA}\ 
and~31.63 for S-H.~Details\ are\ presented\
in\ Table~\ref{tab1}.\ From the computational\ 
results,\ the value for dehydrated system is\ 
higher than the hydrated.\ The MSD for dehydrated\ 
system is greater\ than the hydrated system,~with\ 
a value of $3.49 \times 10^{-4} \hspace{1mm}\mathrm{cm^2 s^{-1}}$\ 
at 333 K.~All O-H pair highest\ peak distances\ 
falls well with\ in the range of a\ hydrogen bond.\ 

\begin{figure}[htbp!]
\centering
\begin{adjustbox}{max size ={\textwidth}{\textheight}}
(\centering a)~\includegraphics[scale=1.0]{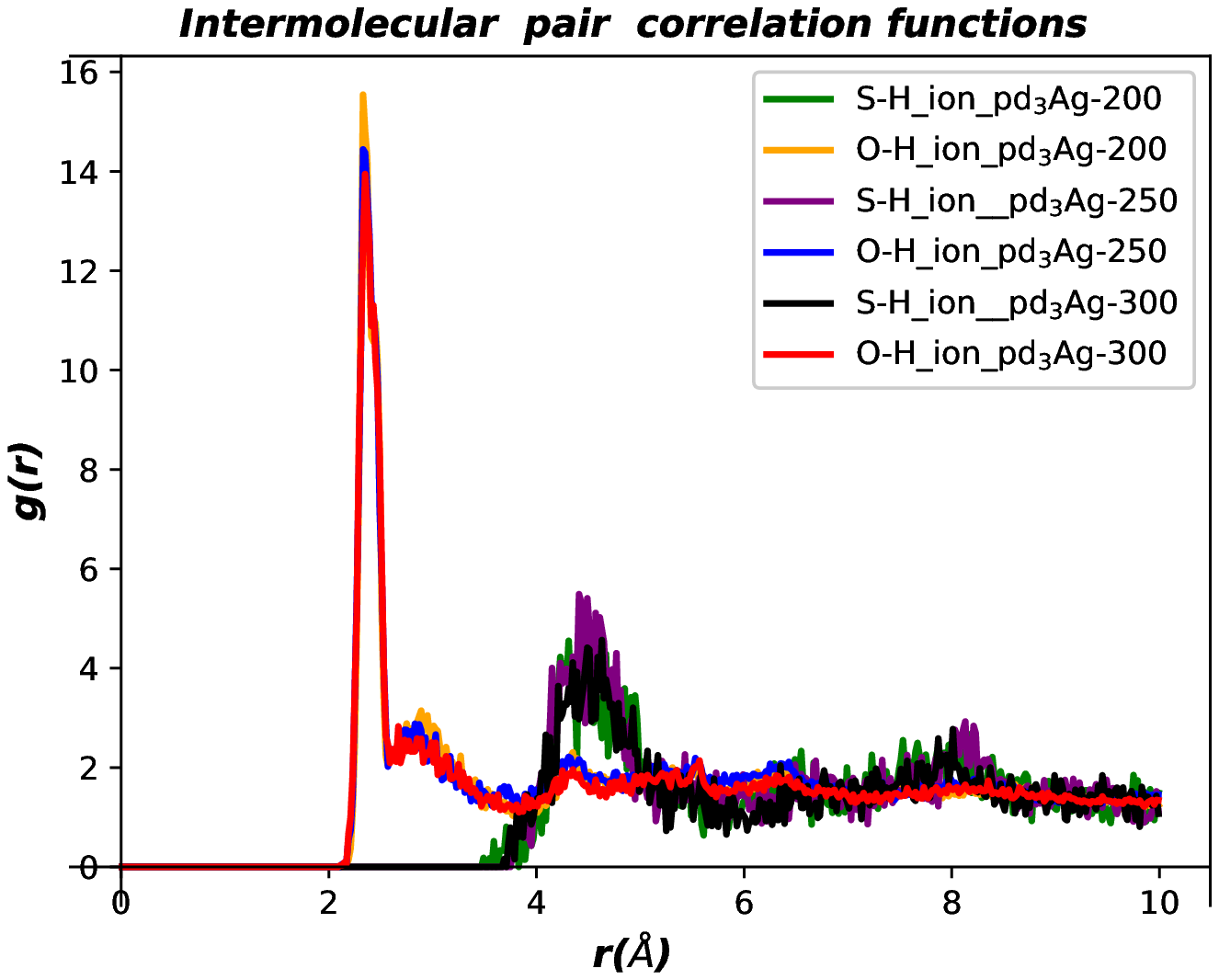}
(\centering b)~\includegraphics[scale=1.0]{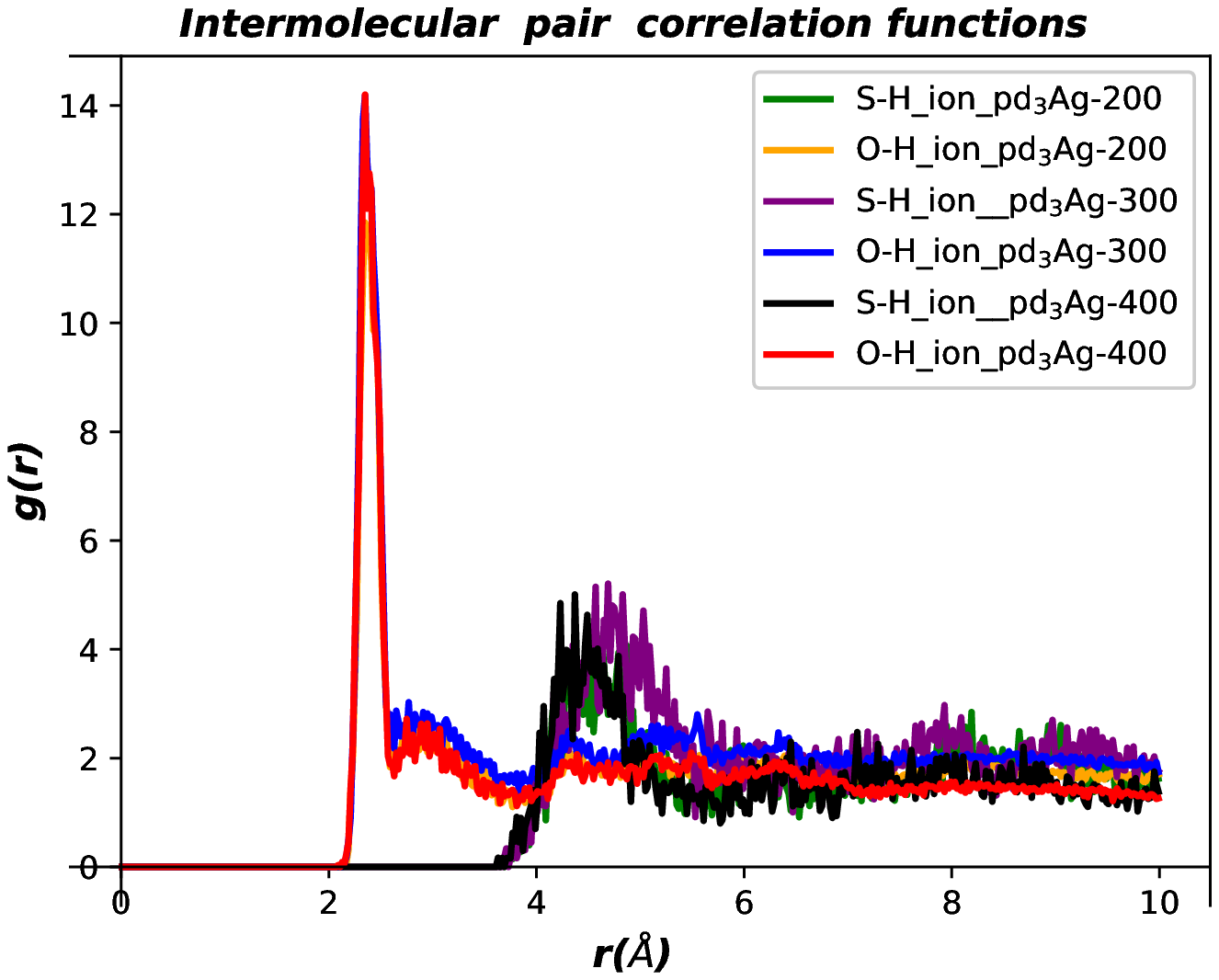}
\end{adjustbox}
\caption{a)~Intermolecular pair correlation functions of O-H\ 
         and S-H at 200 K, 250 K and 300 K with Pd$\rm_{3}$Ag\ 
         electrode.~Here\ the\ number\ of\ protons is 400~(left side).\ 
         b)~Number of protons being 200, 300, and\ 400\ at\ 
         a temperature\ of 333~K~(right side).\label{fig20}}
\end{figure}

The\ intermolecular pair\ correlation functions\ 
of O-H pair,~S-H pair,~at 200 K,~250 K,~and 300 K\ 
are shown\ in Fig.~\ref{fig20}.\ The highest peaks\ 
of O-H pair intermolecular pair correlation functions\ 
occur\ at a positions of~2.33~{\AA},~2.33~{\AA},\ 
and 2.35~{\AA},~for 200 K,~250 K,~and~300 K,~respectively.\ 
The coordination number of O-H pair are 35,~36,~and 34,\ 
respectively,\ for 200 K,~250 K,~and 300 K.~The highest\ 
peaks of S-H pair\ intermolecular pair correlation\ 
functions are\ at a positions of~4.45~{\AA},\ 4.41~{\AA},\ 
and 4.63~{\AA},~respectively,~for 200 K,~250 K,~and\ 
300 K.\ The coordination number\ of S-H pair are\ 
31,\ 32,\ and 30,\ respectively,\ for 200 K,\ 250 K,\ 
and,\ 300 K~(as shown in Table~\ref{tab1}).\ 
The\ fractional coordination\ numbers are results\ 
of a selected\ numerical integrator\ of the Eq.~\eqref{coordinationeq},\
and\ a corresponding\ nearest\ value integers\ 
can be used\ upon\ discussions.\ 

All O-H pair\ highest peak\ distances fall well\ 
within the\ range of a hydrogen\ bond formation\ 
region. 
 %%%%%%%Figure 20................................
The coordination numbers and position of highest peak 
of O-H pair ans S-H pair for different temperatures 
are given in Table~\ref{tab1}.
According to the values of highest peaks falls\ 
in hydrogen bond formation region,\ oxygen\ 
interacts with\ protons by creating hydrogen\ 
bonds,\ but sulfur\ interacts with protons\ 
at a greater\ distance than oxygen.\ The\ 
hydrogen bonding\ network can be\ used\ 
to transfer protons\ from anode to cathode.\

\section{Conclusion\label{sec:conc}}
The molecular dynamics simulation results\ 
suggest\ that the\ conductivity of proton\ 
ions in\ the fuel cell\ decreases as temperature\ 
increases.\ The\ intermolecular pair correlation\ 
functions and\ coordination numbers show that\ 
the sulfonic acid\ group has\ significant\ 
interactions\ and hydrogen bonding.\
The proton mobility\ is aided by hydrogen\ 
bonds forming\ and breaking since\ the highest\ 
peaks of\ O-H are in\ the range\ of hydrogen\ 
bond formation.\ Proton transport\ can be\ 
facilitated by\ a network of\ hydrogen bonds.\ 
When Pt\ is included\ in the\ PEMFCs model,\ 
proton conductivity\ is\ relatively\ higher\ 
compared\ to proton\ conductivity\ when Pd or\ 
Pd$\rm_{3}$Ag is\ used as\ an electrode.\
When Pt/Pd/Pd$\rm_{3}$Ag\ is included\ 
in the\ PEMFCs interaction model,\ proton\ 
conductivity\ is\ improved compared\ to\ 
proton conductivity\ without electrode\ 
contribution.\ The\ conductivity\ values\ 
of\ proton\ ions\ with\ the\ SEEK\ 
electrolyte and Pt electrode\ is\ in\ the\ 
range~[$1.44$~$\times$~10$\rm^{6}$,~$1.85$~$\times$~10$\rm^{6}$]~$S/cm$,\
when\ the\ temperature\ increases\ in\ the\
range~[200,~433]~K.
The\ conductivity\ values\ of\ proton\ 
ions\ with\ the\ SEEK\ electrolyte and Pd electrode\
is\ in\ the\ range~[$1.31$~$\times$~10$\rm^{6}$,~$1.81$~$\times$~10$\rm^{6}$]~$S/cm$,\
when\ the\ temperature\ increases\ in\ the\
range~[200,~433]~K.\ The\ conductivity\ values\ 
of\ proton\ ions\ with\ the\ SEEK\ 
electrolyte~and Pd$\rm_{3}$Ag\ electrode\ is\ 
in\ the\ range~[$7.69$~$\times$~10$\rm^{5}$,~$7.46$~$\times$~10$\rm^{5}$]~$S/cm$,\
when\ the\ temperature\ increases\ in\ the\
range~[200,~433]~K.\ The ion\ transport\ 
properties predicted\ with all potential\ 
electrodes is\ good enough.\ Furthermore,\ 
the ion\ conductivity for\ Pd$\rm_{3}$Ag\ 
electrode of\ PEMFC\ is reduced\ only by\ 
a factor of\ half compared\ to the ion\ 
conductivity\ with either\ of the Pt\ 
electrode PEMFC\ or Pd electrode\ PEMFC.\ 
Thus,\ Pd$\rm_{3}$Ag\ can function\ 
smoothly in replacing the Pt electrode 
PEMFC.\ Furthermore,~dehydrated\
SEEK~electrolyte\ performs\ relatively 
better\ than\ hydrated\ SEEK\ electrolyte.\
\section*{Disclosure\ statement}
The authors declare\ that they have no known\ 
competing financial\ interests or personal\ 
relationships that\ could have appeared\ 
to influence the\ work reported in this paper.\
\section*{Data\ Availability\ Statement}
The data that\ support the findings\ of\ 
this study\ are available\ upon reasonable\ 
request\ from the\ authors.\

\section*{Acknowledgments}
We are grateful to the Ministry of Education\ 
of Ethiopia for financial support.\ The authors\ 
also acknowledge\ the\ Department of Physics at\ 
Addis Ababa\ University.\ 
The~office~of~VPRTT~of Addis\ Ababa\ university\ 
is also warmly~appreciated~for\ supporting~this\ 
research under a\ grant~number~AR/053/2021.\ 
  
\newpage
\section*{References}
\bibliographystyle{elsarticle-num}
\bibliography{refs.bib}

\begin{thebibliography}{10}
\expandafter\ifx\csname url\endcsname\relax
  \def\url#1{\texttt{#1}}\fi
\expandafter\ifx\csname urlprefix\endcsname\relax\def\urlprefix{URL }\fi
\expandafter\ifx\csname href\endcsname\relax
  \def\href#1#2{#2} \def\path#1{#1}\fi

\bibitem{vielstich2009handbook}
W.~Vielstich, H.~A. Gasteiger, H.~Yokokawa,
  \href{https://doi.org/10.1002/cphc.200490023}{Handbook of fuel
  cells:~advances in electrocatalysis, materials, diagnostics and durability},
  Vol.~5, John Wiley $\&$ Sons, 2009.
\newline\urlprefix\url{https://doi.org/10.1002/cphc.200490023}

\bibitem{wieckowski2003catalysis}
A.~Wieckowski, E.~R. Savinova, C.~G. Vayenas,
  \href{https://doi.org/10.1201/9780203912713}{Catalysis and electrocatalysis
  at nanoparticle surfaces}, CRC Press, 2003.
\newline\urlprefix\url{https://doi.org/10.1201/9780203912713}

\bibitem{weber2018carbon}
P.~Weber, D.~J. Weber, M.~Janssen, M.~Werheid, M.~Oezaslan,
  \href{https://doi.org/10.1149/ma2018-02/44/1483}{Carbon supported {Pt-Co}
  {Alloy Nanoparticles} as {HOR} and {ORR} catalyst for {PEM} fuel cells}, in:
  ECS Meeting Abstracts, no.~44, IOP Publishing, 2018, p. 1483.
\newline\urlprefix\url{https://doi.org/10.1149/ma2018-02/44/1483}

\bibitem{antolini2003formation}
E.~Antolini, \href{https://doi.org/10.1016/S0254-0584(02)00389-9}{Formation of
  carbon-supported {PtM} alloys for low temperature fuel cells:~a review},
  Materials chemistry and physics 78~(3) (2003) 563--573.
\newline\urlprefix\url{https://doi.org/10.1016/S0254-0584(02)00389-9}

\bibitem{shao2007understanding}
Y.~Shao, G.~Yin, Y.~Gao,
  \href{https://doi.org/10.1016/j.jpowsour.2007.07.004}{Understanding and
  approaches for the durability issues of {Pt}-based catalysts for {PEM} fuel
  cell}, Journal of Power Sources 171~(2) (2007) 558--566.
\newline\urlprefix\url{https://doi.org/10.1016/j.jpowsour.2007.07.004}

\bibitem{yu2007recent}
X.~Yu, S.~Ye, \href{https://doi.org/10.1016/j.jpowsour.2007.07.049}{Recent
  advances in activity and durability enhancement of {Pt/C} catalytic cathode
  in {PEMFC}:~{Part I}. {Physico-chemical} and electronic interaction between
  {Pt} and carbon support, and activity enhancement of {Pt/C} catalyst},
  Journal of power sources 172~(1) (2007) 133--144.
\newline\urlprefix\url{https://doi.org/10.1016/j.jpowsour.2007.07.049}

\bibitem{chen2009shape}
J.~Chen, B.~Lim, E.~P. Lee, Y.~Xia,
  \href{https://doi.org/10.1016/j.nantod.2008.09.002}{Shape-controlled
  synthesis of platinum nanocrystals for catalytic and electrocatalytic
  applications}, Nano Today 4~(1) (2009) 81--95.
\newline\urlprefix\url{https://doi.org/10.1016/j.nantod.2008.09.002}

\bibitem{peng2009designer}
Z.~Peng, H.~Yang, \href{https://doi.org/10.1016/j.nantod.2008.10.010}{Designer
  platinum nanoparticles:~control of shape, composition in alloy, nanostructure
  and electrocatalytic property}, Nano today 4~(2) (2009) 143--164.
\newline\urlprefix\url{https://doi.org/10.1016/j.nantod.2008.10.010}

\bibitem{liu2011tuning}
Z.~Liu, G.~S. Jackson, B.~W. Eichhorn,
  \href{https://doi.org/10.1039/C1EE01125A}{Tuning the {CO}-tolerance of
  {Pt-Fe} bimetallic nanoparticle electrocatalysts through architectural
  control}, Energy $\&$ Environmental Science 4~(5) (2011) 1900--1903.
\newline\urlprefix\url{https://doi.org/10.1039/C1EE01125A}

\bibitem{HL69}
L.~Hunt, F.~Lever,
  \href{https://doi.org/10.1021/ba-1971-0098.ch005}{Availability of the
  {Platinum Metals}}, Platinum Metals Rev. 13~(4) (1971) 126.
\newline\urlprefix\url{https://doi.org/10.1021/ba-1971-0098.ch005}

\bibitem{shao2006palladium}
M.~Shao, T.~Huang, P.~Liu, J.~Zhang, K.~Sasaki, M.~Vukmirovic, R.~Adzic,
  \href{https://pubs.acs.org/doi/10.1021/la0610553}{Palladium monolayer and
  palladium alloy electrocatalysts for oxygen reduction}, Langmuir 22~(25)
  (2006) 10409--10415.
\newline\urlprefix\url{https://pubs.acs.org/doi/10.1021/la0610553}

\bibitem{Nigussa_2019}
K.~N. Nigussa, \href{https://doi.org/10.1088/2053-1591/ab3fd4}{A study of
  properties of palladium metal as a component of fuel cells}, Materials
  Research Express 6~(10) (2019) 105540.
\newline\urlprefix\url{https://doi.org/10.1088/2053-1591/ab3fd4}

\bibitem{awulachew2022principles}
S.~S. Awulachew, K.~N. Nigussa, {First principles and Monte Carlo studies of
  adsorption and desorption properties from Pd$\rm_{1-x}$Ag$\rm_{x}$ surface
  alloy} (2022).
\newblock \href {http://arxiv.org/abs/2204.02812} {\path{arXiv:2204.02812}}.

\bibitem{sun1998compass}
H.~Sun, \href{https://doi.org/10.1021/jp980939v}{{COMPASS}:~an ab initio
  force-field optimized for condensed-phase applications overview with details
  on alkane and benzene compounds}, The Journal of Physical Chemistry B
  102~(38) (1998) 7338--7364.
\newline\urlprefix\url{https://doi.org/10.1021/jp980939v}

\bibitem{nymand2000ewald}
T.~M. Nymand, P.~Linse, \href{https://doi.org/10.1063/1.481216}{Ewald summation
  and reaction field methods for potentials with atomic charges, dipoles, and
  polarizabilities}, The Journal of Chemical Physics 112~(14) (2000)
  6152--6160.
\newline\urlprefix\url{https://doi.org/10.1063/1.481216}

\bibitem{ibrahim2020go}
A.~Ibrahim, O.~Hossain, J.~Chaggar, R.~Steinberger-Wilckens, A.~El-Kharouf,
  \href{https://doi.org/10.1016/j.ijhydene.2019.05.210}{Go-nafion composite
  membrane development for enabling intermediate temperature operation of
  polymer electrolyte fuel cell}, International Journal of Hydrogen Energy
  45~(8) (2020) 5526--5534.
\newline\urlprefix\url{https://doi.org/10.1016/j.ijhydene.2019.05.210}

\bibitem{cummings1991nonequilibrium}
P.~Cummings, B.~Wang, D.~Evans, K.~Fraser,
  \href{https://doi.org/10.1063/1.459886}{Nonequilibrium molecular dynamics
  calculation of self-diffusion in a non-{Newtonian} fluid subject to a
  {Couette} strain field}, The Journal of chemical physics 94~(3) (1991)
  2149--2158.
\newline\urlprefix\url{https://doi.org/10.1063/1.459886}

\bibitem{allen1987analyse}
M.~P. Allen, D.~J. Tildesley, How to analyse the results, in: Computer
  Simulation of Liquids, Oxford University Press, 1987.

\end{thebibliography}
\end{document}